\documentclass[apjl]{emulateapj}
\usepackage{apjfonts}
\begin{document}
\title{The Super-Alfv\'{e}nic Model of Molecular Clouds:\\
Predictions for Zeeman Splitting Measurements}

\author{Tuomas Lunttila\altaffilmark{1}, Paolo Padoan\altaffilmark{2}, Mika Juvela\altaffilmark{1}, and \AA ke Nordlund\altaffilmark{3}}
\altaffiltext{1}{Helsinki University Observatory, P.O. Box 14, T\"{a}htitorninm\"{a}ki, FI-00014,
University of Helsinki, Finland.}
\altaffiltext{2}{Department of Physics, University of California, San Diego, La Jolla, CA
92093-0424; ppadoan@ucsd.edu.}
\altaffiltext{3}{Astronomical Observatory/Niels Bohr Institute, Juliane Maries Vej 30, DK-2100,
Copenhagen, Denmark.}

\begin{abstract}

We present synthetic OH Zeeman splitting measurements of a super-Alfv\'{e}nic molecular cloud model. We select dense cores from synthetic $^{13}$CO maps computed from the largest simulation to date of supersonic and super-Alfv\'{e}nic turbulence. The synthetic Zeeman splitting measurements in the cores yield a relation between the magnetic field strength, $B$, and the column density, $N$, in good agreement with the observations. The large scatter in $B$ at a fixed value of $N$ is partly due to intrinsic variations in the magnetic field strength from core to core. We also compute the relative mass-to-flux ratio between the center of the cores and their envelopes, ${\cal R}_{\mu}$, and show that super-Alfv\'{e}nic turbulence produces a significant scatter also in ${\cal R}_{\mu}$, including negative values (field reversal between core center and envelope). We find ${\cal R}_{\mu} < 1$ for 70\% of the cores, and ${\cal R}_{\mu} < 0$ for 12\%. Of the cores with $|B_{\rm LOS}|>10$~$\mu$G, 81\% have ${\cal R}_{\mu}<1$. These predictions of the super-Alfv\'{e}nic model are in stark contrast to the ambipolar drift model of core formation, where only ${\cal R}_{\mu}>1$ is allowed.

\end{abstract}
\section{Introduction}
The process of star formation in molecular clouds involves a complex interaction of turbulent velocity fields, gravitational forces, and magnetic fields. While the role of turbulence has been investigated in depth only in recent years, the importance of the magnetic field in star formation was recognized long ago. On small scales, the magnetic field may support sub-critical prestellar cores against gravitational collapse for an ambipolar drift time \citep{Lizano+Shu89}, and provide a way to shed angular momentum by magnetic breaking \citep{Mouschovias77,Mouschovias79} and by helping to power winds and jets \citep{Pudritz+Norman83}. On larger scales, a strong enough magnetic field may even prevent the collapse of a giant molecular cloud (GMC), potentially explaining the low star-formation rate in the Galaxy \citep{Shu+87}.

As an alternative to this scenario of large scale magnetic support, \citet{Padoan+Nordlund99MHD} proposed a super-Alfv\'{e}nic model of molecular clouds. In this model, the turbulent flows can compress the gas in all directions through shocks. Compressions with a component perpendicular to the magnetic field can locally increase the field strength, and the net result is a correlation of magnetic field and gas density, though with a large scatter. While in this model the mean magnetic field of a GMC is very low, its value inside dense cores and filaments (where it is usually observed) can be very large. Based on a number of tests comparing simulations with observations \citep{Padoan+Nordlund99MHD,Padoan+04power}, the super-Alfv\'{e}nic model seems to explain the observations better than models assuming a strong magnetic field.

The most direct way to infer the magnetic field strength in GMCs is by Zeeman splitting measurements. These measurements are very difficult, and are available in GMCs only for relatively dense cores, primarily from OH emission lines \citep[e.g.][]{Crutcher+93,Troland+Crutcher08}. Only a handful of these measurements have provided detections of the field strength. Zeeman splitting is only sensitive to the line-of-sight component of the magnetic field averaged along the line of sight, and weighted by the OH emission (roughly speaking gas density and OH abundance). Thus, Zeeman splitting does not directly provide the large-scale volume-averaged magnetic field strength, needed to understand the nature of the fragmentation process that leads to star formation.

In this Letter, we relate the large-scale mean magnetic field to local Zeeman splitting measurements by a {\it deductive method}, predicting the observational properties of the theoretical model, rather than inducing the theory from the observations. The super-Alfv\'{e}nic molecular cloud model is obtained from a numerical experiment of three-dimensional, supersonic, magneto-hydrodynamic (MHD) turbulence. Synthetic Zeeman splitting measurements are then simulated through radiative transfer calculations. This work is partly motivated by i) an improved interpretation of existing Zeeman splitting measurements by Crutcher (2008, pers. comm.), and ii) a new observational test of ambipolar drift by \citet{Crutcher+08b}.

\begin{figure*}[t]
\epsscale{1.15}
\plottwo{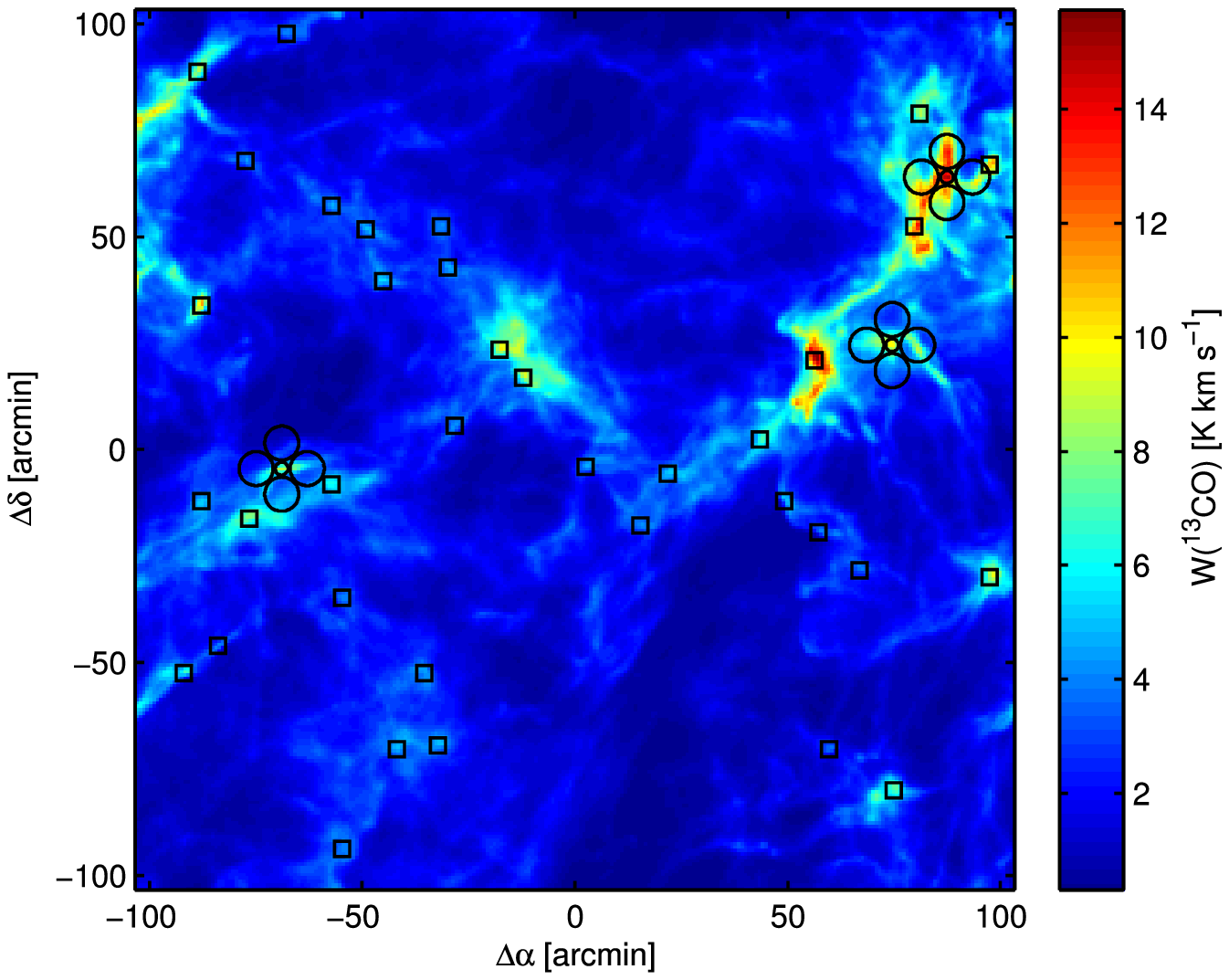}{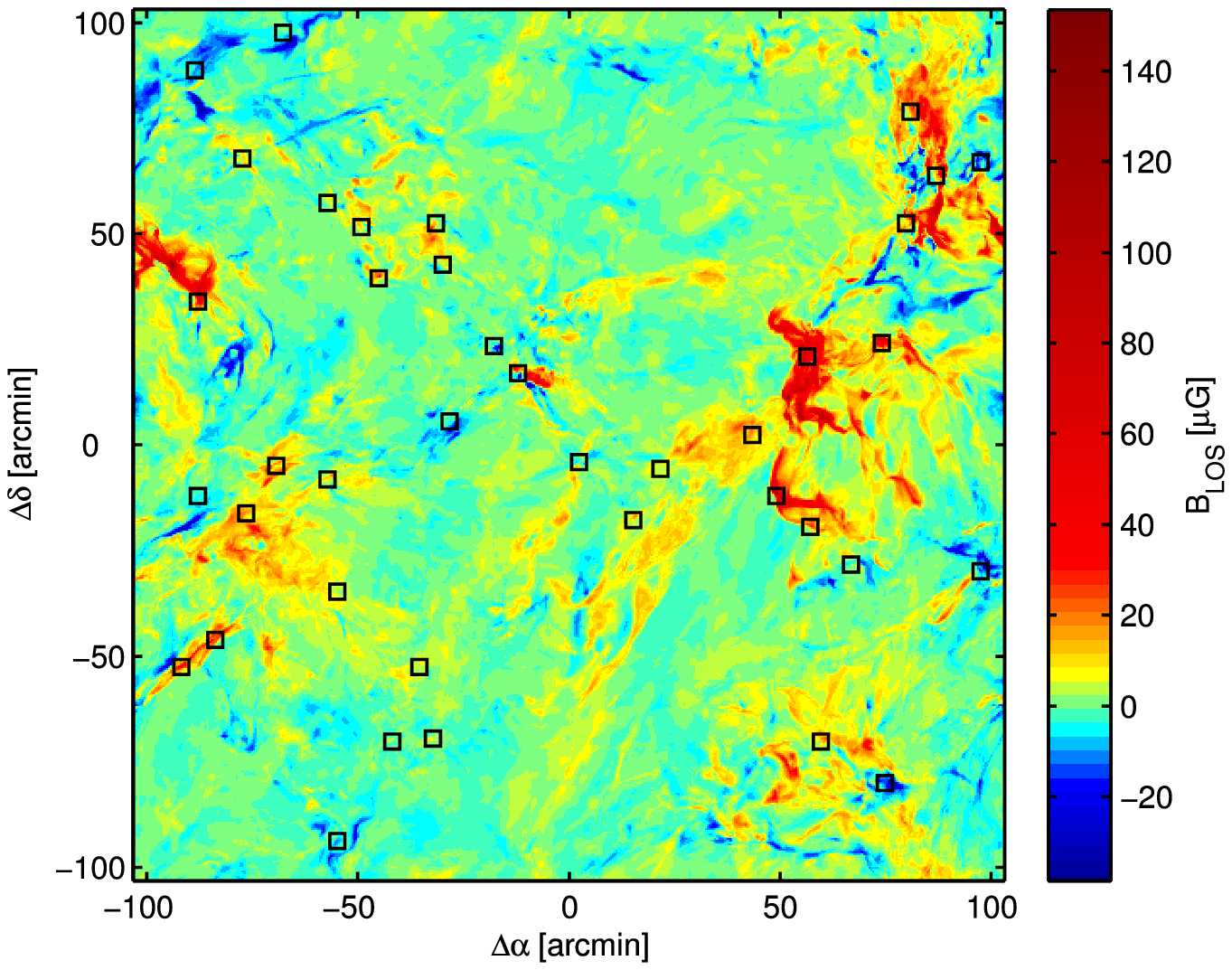}
\caption{\emph{Left:} Simulated $^{13}$CO (1-0) map of the model in the $z$-axis direction. The locations of the cloud cores are shown with squares. The circles indicate the locations of telescope beams used in the synthetic observations of three cores. \emph{Right:} Line-of-sight magnetic field strength as calculated from Zeeman splitting.}
\end{figure*}

A conclusion of the new analysis of Zeeman splitting measurements by Crutcher (2008, pers. comm.) is that the observed scatter in the relation between the absolute value of the line-of-sight magnetic field strength, $| B_{\rm LOS}|$, and the column density, $N$, is partly due to a large scatter in the intrinsic magnetic field strength, $B$, at any given $N$. As shown before \citep{Padoan+Nordlund99MHD}, and as illustrated in the present work, this is a fundamental prediction of the super-Alfv\'{e}nic model. If the mean field were very strong, its local value could not have a large scatter, and the observed scatter in $| B_{\rm LOS}|$ could only arise as a result of random orientations, which seems to be ruled out by the new analysis of Crutcher (2008, pers. comm.).

The new observational test of ambipolar drift by \citet{Crutcher+08b}, is based on the mass to flux ratio of the envelope relative to that of the core -- henceforth referred to as the "relative mass-to-flux ratio", ${\cal R}_{\mu}$, as the ambipolar drift model of core formation predicts ${\cal R}_{\mu}>1$ (e.g. \citet{Ciolek+Mouschovias94} -Fig.~3a). In this Letter we show that the super-Alfv\'{e}nic model predicts a significant scatter in ${\cal R}_{\mu}$ (including negative values), with ${\cal R}_{\mu}<1$ in 81\% of the cores with $| B_{\rm LOS}|>10$~$\mu$G. Values of ${\cal R}_{\mu}<1$ in cores formed by turbulent flows were first found by \cite{Vazquez-Semadeni+05}, where a plausible mechanism for explaining that result was proposed.

\section{Numerical Simulation of Super-Alfv\'{e}nic turbulence}
This work is based on a single snapshot of a supersonic and super-Alfv\'{e}nic ideal MHD turbulence simulation, run on a mesh of $1000^3$ zones with the Stagger Code \citep{Padoan+07imf}. We adopt periodic boundary conditions, isothermal equation of state, random forcing in Fourier space at wavenumbers $1\le k\le 2$ ($k=1$ corresponds to the computational box size), uniform initial density and magnetic field, random initial velocity field with power only at wavenumbers $1 \le k\le 2$.  The rms sonic Mach number of the snapshot used in this work is ${\cal M}_{\rm s}=\sigma_{\rm v,3D}/c_{\rm s}= 8.91$.

The initial magnetic field is such that the initial value of the ratio of gas to magnetic pressure is $\beta_{\rm i}=22.2$. At the time corresponding to the snapshot used for this work, the rms magnetic field strength has been amplified by the turbulence, and the value of $\beta$ defined with the rms magnetic pressure is $\beta=0.2$. This corresponds to an rms Alfv\'{e}nic Mach number of ${\cal M}_{\rm a}=(\beta/2)^{1/2}\sigma_{\rm v,3D} / c_{\rm s}=2.8$, so the turbulence is still super-Alfv\'{e}nic even with respect to the rms Alfv\'{e}n velocity. With respect to the Alfv\'{e}n velocity corresponding to the mean magnetic field ($\beta_{\rm i}=22.2$), the rms Alfv\'{e}nic Mach number is much larger, ${\cal M}_{\rm a,i}=(\beta_{\rm i}/2)^{1/2}\sigma_{\rm v,3D} / c_{\rm s}=29.7$.

\begin{figure*}[t]
\epsscale{1.1}
\plottwo{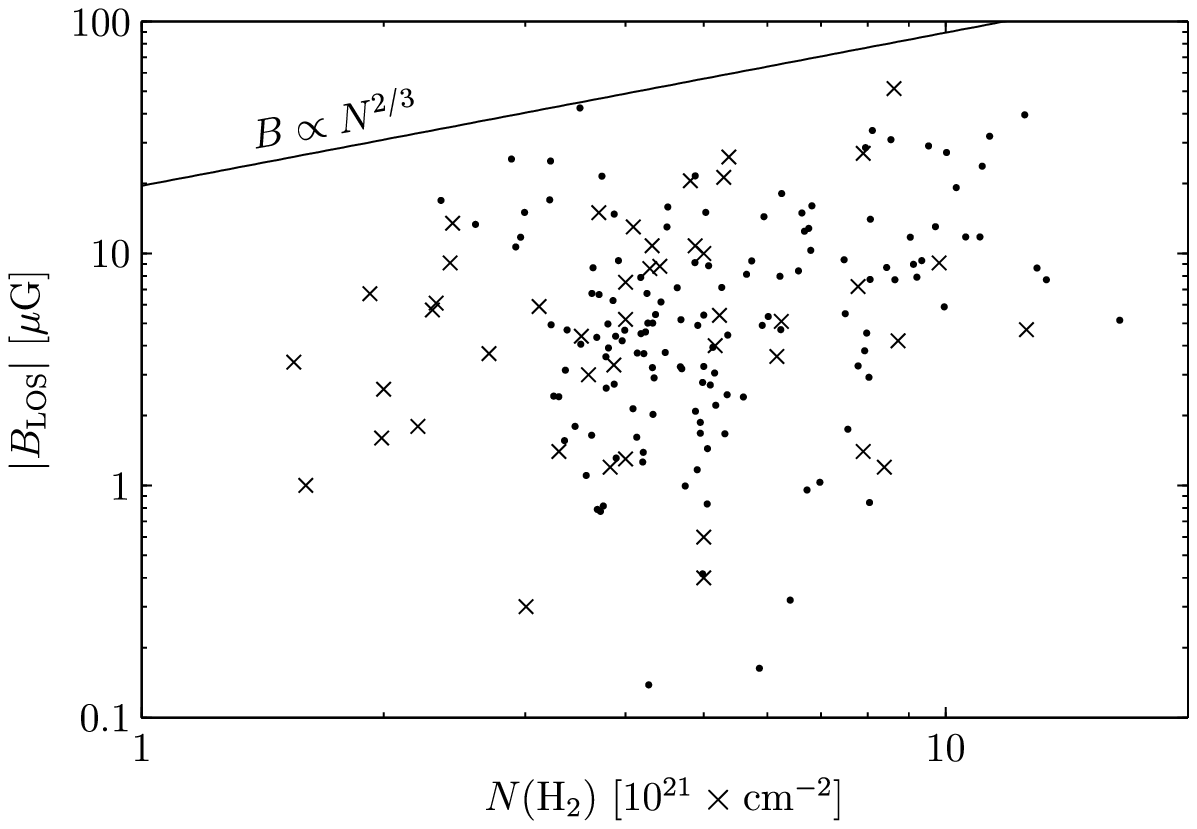}{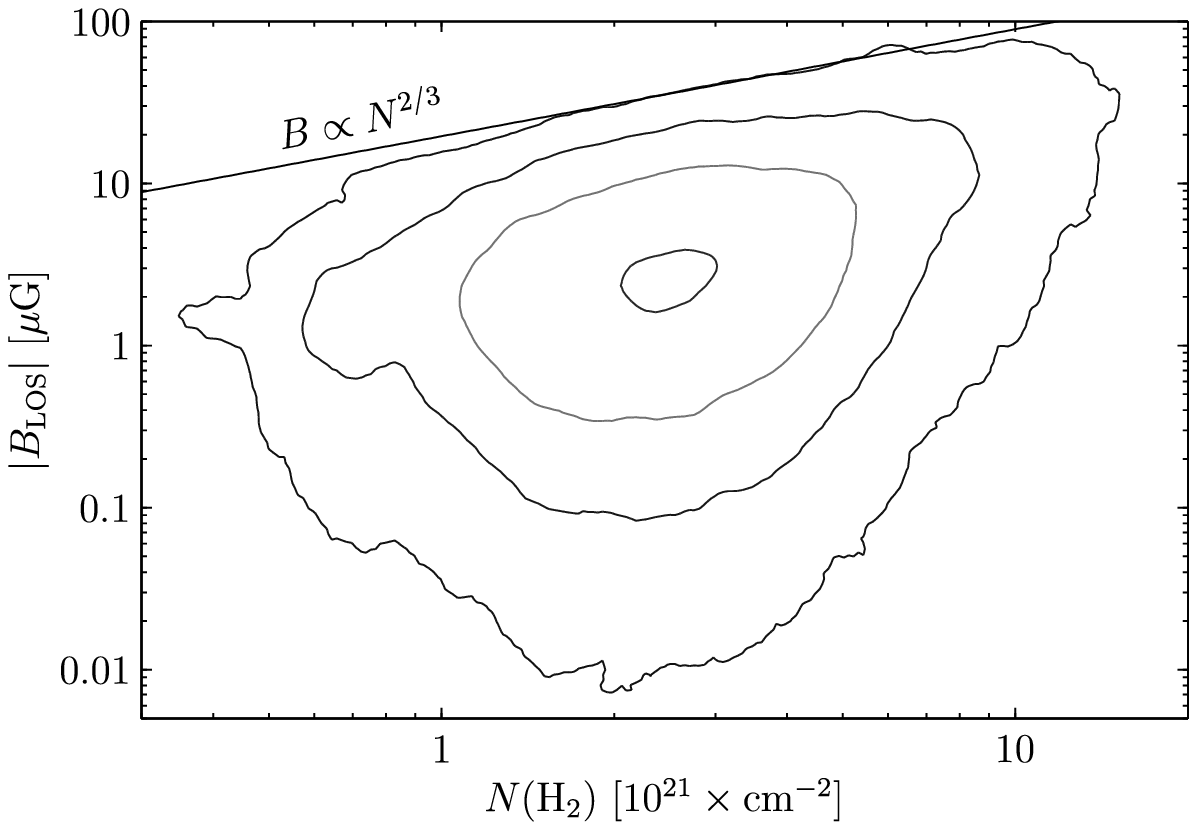}
\caption{\emph{Left:} Estimated line-of-sight magnetic field strength as a function of hydrogen column density. The dots show the results from our simulated observations. The crosses show the results from observations listed in \citet{Troland+Crutcher08} and \citet{Crutcher99}. \emph{Right:} The distribution function for $(N,B)$ from the simulations, using all the points in the three synthetic Zeeman splitting maps. The contours are at levels 0.01, 0.1, 0.4, and 1.5 $(\log_{10}(N)\log_{10}(B))^{-1}$.}
\end{figure*}

\section{Radiative Transfer and Zeeman Splitting Measurements}
For the computation of synthetic Zeeman spectra the data cube is scaled to physical units. The length of the grid is fixed to $L=6$ pc, the mean density to $\left<n(\mathrm{H}_2)\right>=150$~cm$^{-3}$, and the kinetic temperature to $T_{\mathrm{kin}}=20$~K. The mean magnetic field is $(B_x,B_y,B_z)=(0.0, 0.0, 0.69)$~$\mu$G and the rms field $B_{\mathrm{RMS}}=6.45$~$\mu$G. We assume a constant fractional OH abundance of $\mathrm{[OH]/[H]}=4.0\times 10^{-8}$ recommended by \citet{Crutcher79}. To compare the results with observations, the distance to the cloud is fixed to 100~pc, yielding a total angular size of $\sim 3.4\degr$. The calculations are done for three orthogonal directions of a single snapshot.

To simulate the selection of dense cores, we use our line radiative transfer program \citep{Juvela97} to calculate $^{13}$CO~$(1-0)$ maps with an angular resolution of $50\arcsec$, after reducing the model resolution to $256^3$ cells. Cores are handpicked in the $^{13}$CO maps for the Zeeman splitting analysis (the same analysis on a set of cores selected with the Clumpfind algorithm \citep{Williams+94} yields practically the same results). To limit the contamination of envelope observations by emission from nearby structures unrelated to the core, we select only cores separated from each other by at least $10\arcmin$. Some envelopes can still be contaminated by weaker cores that were not selected for the analysis, but in calculations where we ignore the part of the envelope that is contaminated by weak cores we find no significant difference in the final results. We will discuss the core selection and the envelope contamination in detail in a forthcoming paper.

The simulated OH Zeeman splitting observations are obtained with the same set of beams as used in the observations of \citet{Crutcher+08b}. The center of each core, i.e. the position with the highest intensity in the CO map, is observed with a $3\arcmin$ (FWHM) beam, corresponding to measurements with the Arecibo telescope. Each of the envelopes of the cores is observed with four $8\arcmin$ beams, centered $6\arcmin$ north, south, east, and west of the core, simulating observations with the Green Bank Telescope. In the Zeeman splitting calculations we use a resolution corresponding to $512^3$ computational cells.

In our Zeeman splitting analysis we focus on the 1665.40184 MHz OH line (hereafter the 1665 MHz OH line). The OH level populations are estimated assuming that the cloud is optically thin. The radiation field is approximately constant throughout the cloud, and the level populations depend only on the local density. Full radiative transfer calculations with lower resolution models showed that the errors in the excitation temperature of the 1665 MHz OH transition are $\la 1$~K. The coupled radiative transfer equations for the four Stokes parameters are integrated along the line of sight, taking into account the effect of the magnetic field on emission and absorption, to obtain the observed I- and V-spectra. We use a bandwidth of 20~km~s$^{-1}$ with 400 velocity channels. No noise is added to the synthetic spectra.

We determine the line-of-sight magnetic field from the simulated observations by least-squares fitting the numerical derivative of the Stokes I-spectrum to the Stokes V-spectrum, as it is usually done with actual Zeeman splitting observations \citep[e.g.][]{Crutcher+93, Bourke+2001}. Column densities of the cores are estimated from simulated OH spectra. Assuming that the line is optically thin, the column density of OH is obtained as
\begin{equation}
N(\mathrm{OH})[\mathrm{cm}^{-2}]=\frac{4.04\times 10^{14}}{1-T_{\mathrm{bg}}/T_{\mathrm{ex}}}W\quad\mathrm{[K\,km\,s^{-1}]}
\end{equation}
where $W$ is the integrated 1665 MHz OH line area, $T_{\mathrm{bg}}$ is the background continuum brightness temperature (2.73 K), and $T_{\mathrm{ex}}$ is the excitation temperature of the transition \citep{Goss68}. We use $T_{\mathrm{ex}}=20$ K in the analysis of our synthetic observations.
The left panel of Fig.~1 shows the simulated CO map and the locations of the selected cores in the $z$-axis direction. The right panel shows the corresponding line-of-sight magnetic field as determined from the Zeeman splitting.

\begin{figure*}[t]
\epsscale{1.1}
\plottwo{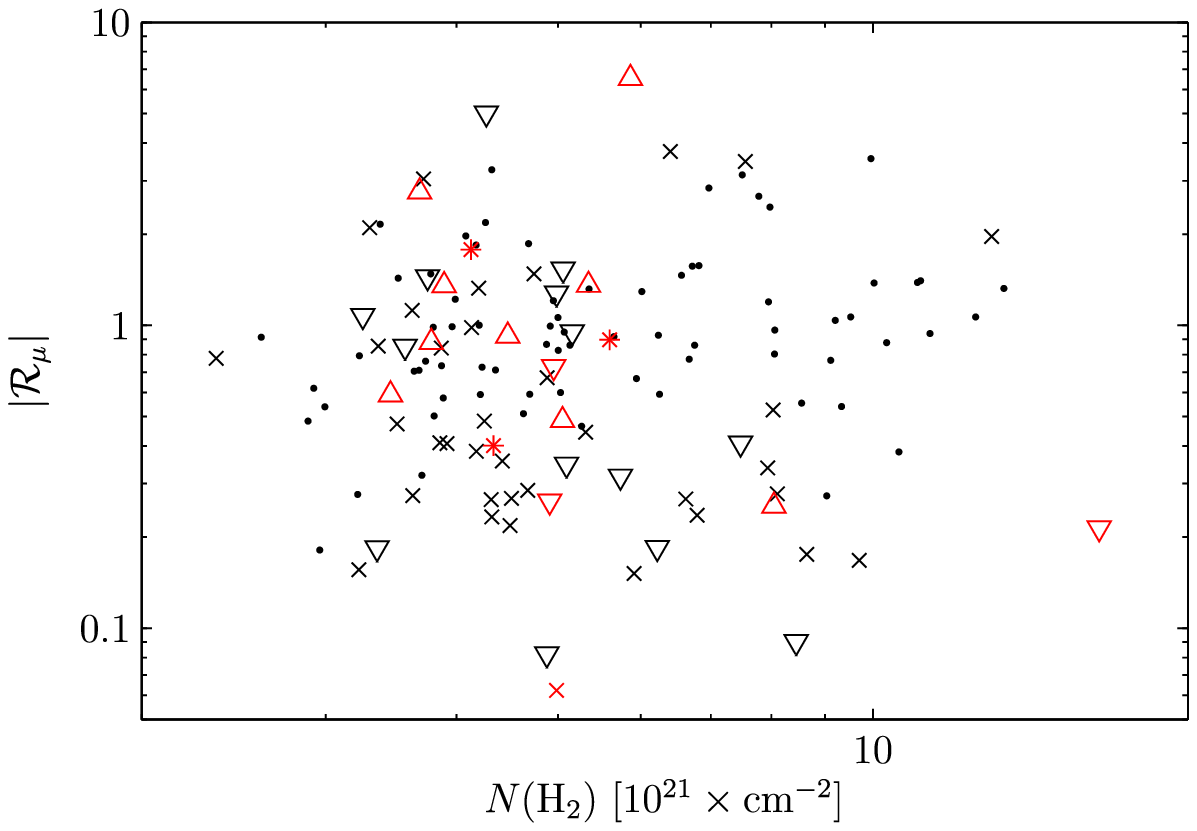}{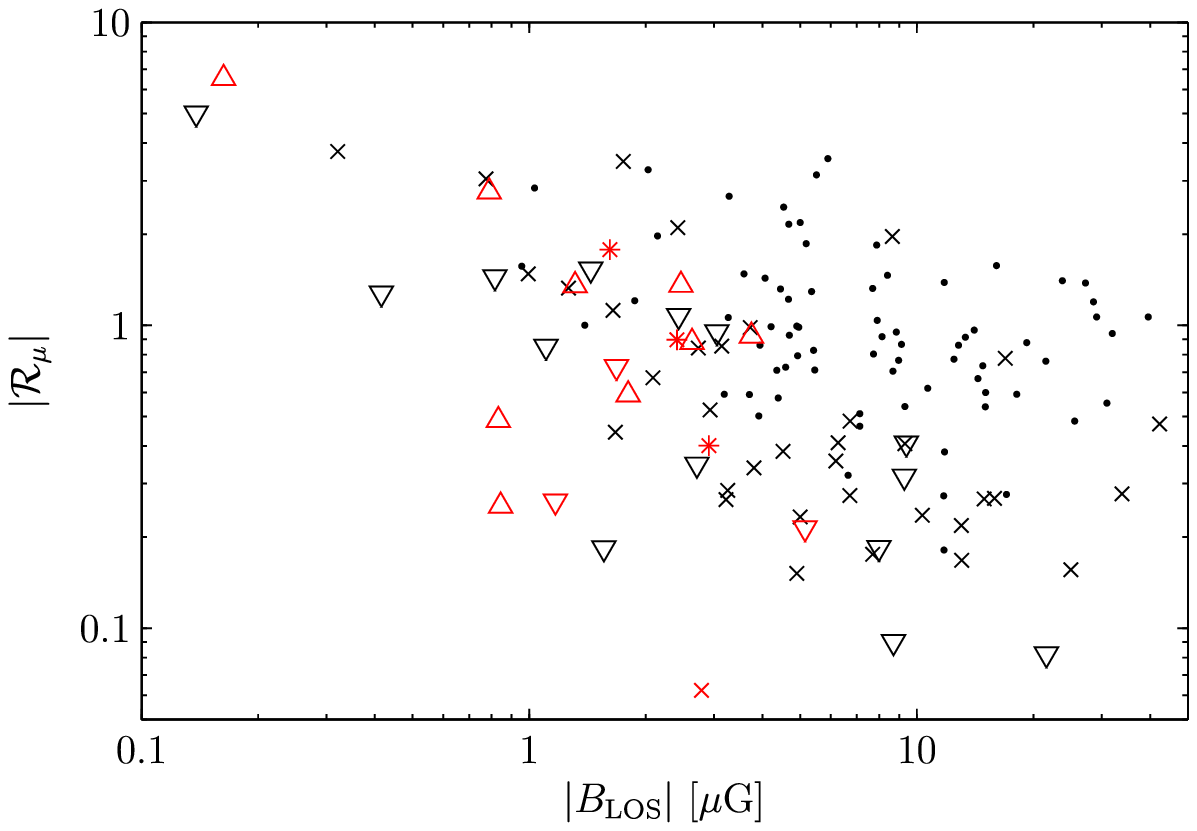}
\caption{\emph{Left:} Relative mass-to-flux ratio for the selected cores as a function of column density. The red symbols indicate the cores with ${\cal R}_{\mu}<0$. The dots, crosses, triangles pointing down, triangles pointing up, and asterisks denote zero, one, two, three, or four field reversals in the envelope relative to the core center, respectively. \emph{Right:} Relative mass-to-flux ratio as a function of inferred magnetic field strength in the central beam. The symbols have the same meaning as in the left panel.}
\end{figure*}

\section{Results}
A total of 139 cores were selected from the $^{13}$CO maps. The left panel of Fig~2 shows the absolute value of their line-of-sight magnetic field strength, $| B_{\mathrm{LOS}}|$, as a function of their H$_2$ column density, $N(\mathrm{H}_2)$. Both the magnetic field and the column density are estimated from the simulated 1665 MHz OH observations with the $3\arcmin$ beam centered on the core. The figure also shows the results of 18 cm OH emission line Zeeman splitting observations from 45 cores that have been reported in the literature. The observations include 34 cores from the recent survey of \citet{Troland+Crutcher08}, and the 11 cores from the list in \citet{Crutcher99} that were observed in 18 cm OH emission lines, and were not included in \citet{Troland+Crutcher08}. The observations and our simulations show a similar distribution of points in the $(N,|B_{\mathrm{LOS}}|)$ space. For a given column density, the inferred line-of-sight magnetic fields are widely scattered. This scatter in the super-Alfv\'{e}nic model is partly due to intrinsic variations in the field strength from core to core, not only to the random orientation of the field. The new analysis of Crutcher (2008, pers. comm.) finds that the distribution of observed $| B_{\mathrm{LOS}}|$ values implies a large intrinsic scatter of the field strength, as in the super-Alfv\'{e}nic model.

The upper envelope of the $| B_{\mathrm{LOS}}|$-$N(\mathrm{H}_2)$ scatter plot is a useful constraint to compare the simulations with the observations. However, given the small number of cores and their narrow range in column density, this upper envelope is not well defined. On the right panel of Fig.~2 we show the contour plot of $| B_{\mathrm{LOS}}|$ versus $N(\mathrm{H}_2)$ for all the points in the three synthetic Zeeman splitting maps. In this case the upper envelope is well defined, and approximately consistent with $| B_{\mathrm{LOS}}| \propto N(\mathrm{H}_2)^{2/3}$. \citet{Padoan+Nordlund99MHD} found a similar upper envelope in the correlation between the three-dimensional distributions of magnetic field strength and gas density, $B \propto n^{0.4}$. This smaller exponent is consistent with the current result, because the gas density spans a much wider range of values than the column density. The power-law upper envelope derived from the right panel of Fig.~2 is replicated on the left panel, showing that it may be approximately valid also for the cores.

Our simulation matches well the observed values of $B_{\mathrm{LOS}}$ in dense cores, despite the very low value of the mean field, $ \langle B \rangle=0.69$~$\mu$G parallel to the $z$ axis, and $\langle B \rangle=0.0$~$\mu$G in the other two orthogonal directions. If we compute the mean field by averaging over the Zeeman splitting maps, we obtain $\langle B_{\mathrm{LOS}} \rangle=2.1$~$\mu$G in the $z$-axis direction ($\sim 0.1$~$\mu$G in the other two directions). Therefore, if one could ideally detect the Zeeman splitting everywhere on a cloud map like our synthetic map in the $z$-axis direction, the value of $\langle B_{\mathrm{LOS}} \rangle $, averaged over the whole map, would be 2.1~$\mu$G, while the correct average value is only  0.69~$\mu$G. More realistically, Zeeman splitting would be detected only in a few dense cores, and the mean value of those detections would overestimate the large-scale mean magnetic field by an even larger factor (by more than one order of magnitude in the case of the cores we have selected in the direction of the $z$ axis). The volume average magnetic field is always overestimated by an average of Zeeman splitting detections because the measurements are essentially density weighted, and the magnetic field tends to be stronger at larger density, as a result of the $B-n$ correlation responsible for the upper envelope of the $B_{\mathrm{LOS}}-N$ scatter plot. Furthermore, only relatively large magnetic fields can be detected, biasing the average of Zeeman splitting detections toward even larger values.

\citet{Crutcher+08b} propose a new method for testing the ambipolar drift model of core formation. They compare the mass-to-flux ratio between the center of a core and its envelope, by measuring the relative mass-to-flux ratio, ${\cal R}_{\mu}$, defined as
\begin{equation}
{\cal R}_{\mu}=\frac{[N(\mathrm{H}_2)/B_{\mathrm{LOS}}]_{\mathrm{core}}}{[N(\mathrm{H}_2)/B_{\mathrm{LOS}}]_{\mathrm{envelope}}}=
\frac{[N(\mathrm{OH})/B_{\mathrm{LOS}}]_{\mathrm{core}}}{[N(\mathrm{OH})/B_{\mathrm{LOS}}]_{\mathrm{envelope}}}.
\end{equation}
The relative mass-to-flux ratio provides a direct test of the ambipolar drift model that requires ${\cal R}_{\mu}>1$ (e.g. \citet{Ciolek+Mouschovias94} -Fig.~3a). This test avoids some of the problems faced by the direct measurement of the mass-to-flux ratio in the cores. For instance, it is not necessary to know the relative abundance of the tracer molecule, or the orientation of the magnetic field, although it is assumed that the relative abundance and field direction do not change from the core center to its envelope. We define the envelope $N(\mathrm{OH})$ and $B_{\mathrm{LOS}}$ as the mean of their values from the four envelope beams. Because the Zeeman splitting gives also the sign of $B_{\mathrm{LOS}}$, ${\cal R}_{\mu}$ is a signed quantity, with negative values indicating a reversal of the mean field in the envelope with respect to the central beam. Field reversals are interesting because they are not predicted by the ambipolar drift model.

Figure~3 shows the absolute value of the relative mass-to-flux ratios from our simulation, plotted against the column density (left panel) and the absolute value of the line-of-sight magnetic field in the core center (right panel). The values of $|{\cal R}_{\mu}|$ are widely scattered, with 91\% of the cores in the range $0.2< |{\cal R}_{\mu}|< 5$. While we do not find any clear trend with column density, there seems to be a correlation between  $|{\cal R}_{\mu} |$ and $| B_{\rm LOS}|$, with  $|{\cal R}_{\mu} | \propto | B_{\mathrm{LOS}}|^{-0.26}$. We find ${\cal R}_{\mu} < 1$ for 70\% of the cores, and ${\cal R}_{\mu}< 0$ for 12\%.  Of the cores with $| B_{\rm LOS}|>10$~$\mu$G (values that could be observationally detected), 81\% have ${\cal R}_{\mu}<1$. Even in cores with ${\cal R}_{\mu}>0$, it is possible to find a field reversal in some of the four envelope beams, relative to the central one (i.e., the line-of-sight component of the magnetic field has opposing signs in the two beams). Figure~3 highlights these cores, with different symbols depending on the number of envelope beams with a reversed field direction. If we consider only the 36 cores with central values of $| B_{\rm LOS}|>10$~$\mu$G, as weaker fields are less likely to be observationally detected, we find nine cores with one envelope field reversal, and one core with two reversals.

\section{Conclusions}
We have found that the super-Alfv\'{e}nic model reproduces well the observed $|B_{\rm LOS}|-N$ relation in dense cores, despite its very low mean magnetic field strength. Because the large scatter of $| B_{\rm LOS}|$ in the simulation is partly due to intrinsic variations of $B$ from core to core, and not only to the effect of the random orientation, we interpret the observations to also be consistent with intrinsic variations in $B$ from core to core. This interpretation of the observations, confirmed by the new analysis of Crutcher (2008, pers. comm.), suggests that the mean magnetic field on a larger scale cannot be as large as in the ambipolar drift model.

The super-Alfv\'{e}nic model also predicts that OH Zeeman splitting measurements of cores with central field values of $| B_{\rm LOS}|>10$~$\mu$G should yield ${\cal R}_{\mu}<1$ in more than $4/5$ of the cores, and one or more field reversals in the envelope beams, relative to the central one, in 28\% of the cores. In contrast, the ambipolar drift model of core formation predicts ${\cal R}_{\mu} >1$ and no field reversal. If future observational measurements of ${\cal R}_{\mu}$ \citep{Crutcher+08b} were to yield ${\cal R}_{\mu}<1$ in most cores, as in the super-Alfv\'{e}nic model, the ambipolar drift model of core formation would seem to be ruled out.

\acknowledgements

This research was partially supported by NASA ATP grant NNG 05-6601G, and by NSF grant AST 05-07768. We utilized computing resources provided by the San Diego Supercomputer Center and by NASA High End Computing Program.  TL and MJ acknowledge the financial support of the Academy of Finland grant 124620.

\bibliographystyle{apj}
\bibliography{Zeeman,padoan,MC}

\begin{thebibliography}{20}
\expandafter\ifx\csname natexlab\endcsname\relax\def\natexlab#1{#1}\fi

\bibitem[{{Bourke} {et~al.}(2001){Bourke}, {Myers}, {Robinson}, \&
  {Hyland}}]{Bourke+2001}
{Bourke}, T.~L., {Myers}, P.~C., {Robinson}, G., \& {Hyland}, A.~R. 2001, ApJ,
  554, 916

\bibitem[{{Ciolek} \& {Mouschovias}(1994)}]{Ciolek+Mouschovias94}
{Ciolek}, G.~E., \& {Mouschovias}, T.~C. 1994, \apj, 425, 142

\bibitem[{{Crutcher}(1979)}]{Crutcher79}
{Crutcher}, R.~M. 1979, \apj, 234, 881

\bibitem[{{Crutcher}(1999)}]{Crutcher99}
---. 1999, ApJ, 520, 706

\bibitem[{{Crutcher} {et~al.}(2008){Crutcher}, {Hakobian}, \&
  {Troland}}]{Crutcher+08b}
{Crutcher}, R.~M., {Hakobian}, N., \& {Troland}, T.~H. 2008, preprint (astro-ph/0807.2862)

\bibitem[{Crutcher {et~al.}(1993)Crutcher, Troland, Goodman, Heiles, Kaz\`{e}s,
  \& Myers}]{Crutcher+93}
Crutcher, R.~M., Troland, T.~H., Goodman, A.~A., Heiles, C., Kaz\`{e}s, I., \&
  Myers, P.~C. 1993, ApJ, 407, 175

\bibitem[{{Goss}(1968)}]{Goss68}
{Goss}, W.~M. 1968, \apjs, 15, 131

\bibitem[{Juvela(1997)}]{Juvela97}
Juvela, M. 1997, A\& A, 322, 943

\bibitem[{{Lizano} \& {Shu}(1989)}]{Lizano+Shu89}
{Lizano}, S., \& {Shu}, F.~H. 1989, \apj, 342, 834

\bibitem[{{Mouschovias}(1977)}]{Mouschovias77}
{Mouschovias}, T.~C. 1977, \apj, 211, 147

\bibitem[{{Mouschovias}(1979)}]{Mouschovias79}
---. 1979, \apj, 228, 159

\bibitem[{{Padoan} {et~al.}(2004){Padoan}, {Jimenez}, {Juvela}, \&
  {Nordlund}}]{Padoan+04power}
{Padoan}, P., {Jimenez}, R., {Juvela}, M., \& {Nordlund}, {\AA}. 2004, \apjl,
  604, L49

\bibitem[{Padoan \& Nordlund(1999)}]{Padoan+Nordlund99MHD}
Padoan, P., \& Nordlund, {\AA}. 1999, ApJ, 526, 279

\bibitem[{{Padoan} {et~al.}(2007){Padoan}, {Nordlund}, {Kritsuk}, {Norman}, \&
  {Li}}]{Padoan+07imf}
{Padoan}, P., {Nordlund}, {\AA}., {Kritsuk}, A.~G., {Norman}, M.~L., \& {Li},
  P.~S. 2007, \apj, 661, 972

\bibitem[{{Pudritz} \& {Norman}(1983)}]{Pudritz+Norman83}
{Pudritz}, R.~E., \& {Norman}, C.~A. 1983, \apj, 274, 677

\bibitem[{{Shu} {et~al.}(1987){Shu}, {Adams}, \& {Lizano}}]{Shu+87}
{Shu}, F.~H., {Adams}, F.~C., \& {Lizano}, S. 1987, ARA\&A, 25, 23

\bibitem[{{Troland} \& {Crutcher}(2008)}]{Troland+Crutcher08}
{Troland}, T.~H., \& {Crutcher}, R.~M. 2008, \apj, 680, 457

\bibitem[{{V{\'a}zquez-Semadeni} {et~al.}(2005){V{\'a}zquez-Semadeni}, {Kim},
  {Shadmehri}, \& {Ballesteros-Paredes}}]{Vazquez-Semadeni+05}
{V{\'a}zquez-Semadeni}, E., {Kim}, J., {Shadmehri}, M., \&
  {Ballesteros-Paredes}, J. 2005, \apj, 618, 344

\bibitem[{{Williams} {et~al.}(1994){Williams}, {de Geus}, \&
  {Blitz}}]{Williams+94}
{Williams}, J.~P., {de Geus}, E.~J., \& {Blitz}, L. 1994, \apj, 428, 693

\end{thebibliography}
\end{document}